\documentstyle[preprint,aps]{revtex}
\begin{document}
\title{ EXACT EQUATION FOR WILSON LOOPS IN 2-DIMENTIONAL EUCLIDEAN SPACE}
\author{ MOFAZZAL AZAM}
\address{THEORETICAL PHYSICS DIVISION, CENTRAL COMPLEX,
BHABHA ATOMIC RESEARCH CENTRE \\
TROMBAY, MUMBAI-400085, INDIA}
\maketitle
\vskip .8 in
\begin{abstract}
We derive an exact equation for simple self non-intersecting Wilson loops
in non-abelian gauge theories with gauge fields interacting with
fermions in 2-dimensional Euclidean space.

\end{abstract}
\newpage
In the context of quantum gravity, string theory and black
holes, t'Hooft \cite{thooft}
and Susskind \cite{suss} have proposed the holographic principle.
According to this
principle, it may be possible to describe the world in terms of variables
living on a two dimensional surface much like information coded on a
two dimensional hologram.On the hologram, Susskind \cite{suss1}
has argued that
the non-local objects such as Wilson loops are the basic observables.These
observables have come to be
known as Precursors.This, indeed, makes Wilson loops very important
objects of study.Long ago, it was shown by this author that in the
set of all posssible Wilson loops, there exists a smaller set which
generates the whole set of observables \cite{azam1}.
It was also shown that, when the
potential and field strenghth are analytic, there exist gauge invariant
local generators for the Wilson loops \cite{azam2}.This last result,
as pointed out by Giddings and Lippert \cite{gid}, may possibly be
relevant while connecting local events deep inside 3-space with Wilson
loops on the hologram.
\par There are  many other important works associated with the Wilson
loops.Notable among them are the works of Polyakov \cite{pol},
Migdal and Makeenko \cite{mig} ,
Jeviki and Sakita  \cite{jev} .Polyakov poined out
that at the classical level the equations
of motion of the Wilson loops in pure gauge theory are very much like
classical equations of motion for chiral fields.In the quantum theory,
Migdal and Makeenko
have derived equations of motion for Wilson loops  in a pure gauge theory
in the large N limit .On the other hand,
Jeviki and Sakita have tried to formulate
the quantum theory in terms of Wilson loops by making a change of variable
in the functional integral for gauge fields.Recently, Agarwal and Rajeev
have provided a geometrically unified picture of all the three above
mentioned approaches \cite{agar}.

\par As mentioned above, the importance of precursors makes us look at
all these and any other
results connected with Wilson loops very seriously.In this paper, we
will be concerned with Wilson loops in a non-abelian gauge theory
with gauge fields interacting
with fermions in
2-dimensional Euclidean space.In fact, we will derive an exact,
closed equation for simple self non-intersecting  Wilson loops.
\par It turns out that 2-dimensional ( or rather 1+1 dimensional) gauge
theories have been fertile ground for many new and important ideas.
The most significant result is Schwinger's demonstration of mass term for
photons in $1+1$ dimensional  QED \cite{schw}.
For SU($N$) gauge theories t'Hooft has
provided a complete solution in the large $N$ limit \cite{thooft1}.
There has also been
extensive study of this model by Belevedre {\it et al} \cite{bel}.
Another interesting
result is by the present author and Naik \cite{azam3}.
In this publication
it was shown that the order parameter of glueball condensation and
discrete chiral symmetry are related, in other words, glueball
condesation would lead to the breaking of discrete chiral
symmetry.
\par The present work is motivated by an attempt to obtain results
similar to Schwinger's in 2-dimensional QED.(From now on,
we will consider only 2-dimensional Euclidean space).We obtain,
in the non-abelian gauge theory
a closed equation for smeared gauge invariant field srtength.
However, the equation
is non-local and we can not interpret it as an equation for
massive pseudoscalar field.In the case of abelian U(1)
gauge theory, this equation
exactly reproduces Schwinger's result.To see what
we mean, let us rederive Schwinger's result in a very simple
and straight forward manner.The equation of motion is given by,
\begin{eqnarray}
\partial^{\mu} F_{\mu \nu}(x)= ~gJ_{\nu}(x)
\end{eqnarray}
where $F_{\mu \nu}$ is the field strength and
$J_{\nu}(x)=\bar{\psi}(x)\gamma_{\nu}\psi(x)$ , $\bar{\psi}(x)$,
$\psi(x)$ are the Dirac spinors and $\gamma_{\nu}$ are the Dirac
gamma matrices in two Euclidean dimensions.The axial vector current
$J_{\mu}^{5}(x)=\bar{\psi}(x)\gamma_{\mu}\gamma^{5}\psi(x)$, has an
anomaly given by \cite{john,col},
\begin{eqnarray}
\partial^{\mu} J_{\mu}^{5}(x)=\frac{g}{\pi}\epsilon^{\mu \nu}F_{\mu \nu}(x)
\end{eqnarray}
where $\epsilon^{\mu \nu}=-\epsilon^{\nu \mu}$ is antisymmetric tensor
in 2-dimensional space.
Note that in 2-dimensional space ,
$F_{\mu \nu}(x)=\epsilon^{\mu \nu}\phi(x)$, where
$\phi(x)$ is a pseudo-scalar field.In terms of the pseudo-scalar field
$\phi(x)$, the equation of motion and the anomaly equation becomes,
\begin{eqnarray}
\partial_{\mu}\phi(x)= ~gJ^{5}_{\mu}(x) \nonumber\\
\partial_{\mu}J^{5}_{\mu}(x)=\frac{g}{\pi} \phi(x)
\end{eqnarray}
From the equations above we easily obtain,
\begin{eqnarray}
\partial^{\mu}\partial_{\mu}\phi(x) = g \partial^{\mu}J^{5}_{\mu}(x)
=\frac{g^2}{\pi}\phi(x)
\end{eqnarray}
This can also be written as,
\begin{eqnarray}
\Box \phi(x) - \frac{g^2}{\pi}\phi(x)=0
\end{eqnarray}
By analytic continuation of the time co-ordinate
to Minkowskii space, we obtain
the equation for a pseudo-scalar particle of mass
$m=\frac{g}{\sqrt{\pi}}$.This was the result obtained by Schwinger
for two dimensional QED.In a non-abelian gauge theory  with gauge fields
interacting
with fermions, we will try to derive an equation analogous to Eq.(5), which
will turn to be a non-local equation for the area derivative of the
Wilson loop.This area derivative of the Wilson loop is, in fact, the
smeared gauge invariant field strength.
\par First of all, let us introduce a few definitions.Let $SU(N)$ be
the non-abelian gauge group with generators $T_a$.
\begin{eqnarray}
[T_a,T_b]=f_{ab}^{c}T_c
\end{eqnarray}
$f_{ab}^{c}$ are the structure constants of the Lie algebra
of the group $SU(N)$.The potential and the field strength $A_{\mu}(x)$
and $F_{\mu \nu}(x)$ are,
\begin{eqnarray}
A_{\mu}(x)=A_{\mu}^{a}(x)T_a ~~~;~~~ F_{\mu \nu}(x)=F_{\mu \nu}^{a}(x)T_a
\end{eqnarray}
Let $C_{xy}$ be a curve starting at $x$ and ending at $y$.$~w(C_{xy})$
is defined as,
\begin{eqnarray}
w(C_{xy})=P \exp({\int_{C_{xy}}A_{\mu}dx^{\mu}})
\end{eqnarray}
If $C_{xx}$ is a curve starting at $x$ and also ending at $x$,
then,
\begin{eqnarray}
w(C_{xx})=P \exp({\oint_{C_{xx}}A_{\mu}dx^{\mu}})
\end{eqnarray}
Note that under gauge transformation, $w(C_{xx})$ transforms exactly
like the field strength $F_{\mu \nu}(x)$.
The Wilson loop
is given by,
\begin{eqnarray}
w(C)=Tr(P \exp({\oint_{C}A_{\mu}dx^{\mu}}))
\end{eqnarray}
The area derivative of the Wilson loop at the point $x$, is given
by \cite{pol,mig}
\begin{eqnarray}
\frac{\delta w(C)}{\delta \sigma_{\mu \nu}(x)}=Tr[F_{\mu \nu}(x) w(C_{xx})]
\end{eqnarray}
In two dimensional space,
\begin{eqnarray}
F_{\mu \nu}=\epsilon_{\mu \nu} \Phi =\epsilon_{\mu \nu} \Phi^{a}T_a ~~;
~~\sigma_{\mu \nu}=\epsilon_{\mu \nu} \sigma
\end{eqnarray}
where $~\epsilon_{\mu \nu}=-\epsilon_{\nu \mu}~ $ is the antisymmetric
tensor in 2-dimensional space.This allows us to write the
area  derivative of the Wilson loop as,
\begin{eqnarray}
\frac{\delta w(C)}{\delta \sigma (x)}=Tr[\Phi (x) w(C_{xx})]
\end{eqnarray}
The area derivative can be looked upon as smeared
gauge invariant field strength.
Note that for abelian gauge fields,
\begin{eqnarray}
\frac{\delta w(C)}{\delta \sigma_{\mu \nu}(x)}=F_{\mu \nu}(x) w(C_{xx})
=~~w(C_{xx}) F_{\mu \nu}(x)
\end{eqnarray}
The non-local derivative $\partial^{x}_{\mu}$ is defined as \cite{pol,mig},
\par (1) For a local function $f(x)$,
\begin{eqnarray}
\partial^{x}_{\mu} f(x)= \partial_{\mu} f(x)
\end{eqnarray}
(2) For non-local objects such as $~w(C_{xy})$,
\begin{eqnarray}
\partial^{x}_{\mu} w(C_{xy})=-A_{\mu}(x)w(C_{xy})~~;~~
\partial^{y}_{\mu} w(C_{xy})=w(C_{xy})A_{\mu}(y)
\end{eqnarray}
(3) For non-local objects  $w(C_{xx})$, corresponding to closed loops
$C_{xx}$,
\begin{eqnarray}
\partial^{x}_{\mu} w(C_{xx}) = [~w(C_{xx})~,~A_{\mu}(x)~]
\end{eqnarray}
Note that for abelian gauge theories,
\begin{eqnarray}
\partial^{x}_{\mu} w(C_{xx}) = [~w(C_{xx})~,~A_{\mu}(x)~]=0
\end{eqnarray}
In 2-dimensional non-abelian gauge theory, the equation of motion is
given by,
\begin{eqnarray}
D^{\mu} F_{\mu \nu}=~gJ_{\nu}
\end{eqnarray}
which can also be written as,
\begin{eqnarray}
D_{\mu} \Phi  =~g \epsilon_{\mu \nu} J^{\nu}=g J^{5}_{\mu}
\end{eqnarray}
The axial vector current $J^{5}_{\mu}$ has an anomaly given by \cite{bel},
\begin{eqnarray}
D^{\mu} J_{\mu}^{5 a}= \frac{g}{2\pi} \Phi^{a}
\end{eqnarray}
Taking non-local derivative $\partial^{x}_{\mu}$
of both side of Eq.(13), after some
algebraic manipulation, we obtain
\begin{eqnarray}
\partial^{x}_{\mu} \frac{\delta w(C)}{\delta \sigma (x)}
=Tr[D_{\mu} \Phi (x) w(C_{xx})]
\end{eqnarray}
Using the equation of motion, we obatin
\begin{eqnarray}
\partial^{x}_{\mu} \frac{\delta w(C)}{\delta \sigma (x)}
=Tr[D_{\mu} \Phi (x)~ w(C_{xx})] =~g~ Tr[J_{\mu}^{5}~w(C_{xx})]
\end{eqnarray}
Applying again the derivative $\partial^{x}_{\mu}$
to both side of Eq.(23) and using the anomaly equation, Eq.(21) we get
\begin{eqnarray}
\partial^{\mu x}~ \partial^{x}_{\mu} \frac{\delta w(C)}{\delta \sigma (x)}
=~g~ Tr[D^{\mu} J_{\mu}^{5}~w(C_{xx})]
=~\frac{g^2}{2\pi}Tr[\Phi (x) w(C_{xx})]
\end{eqnarray}
Let  $\Box ^{x} \equiv
\partial^{\mu x}~ \partial^{x}_{\mu}$.Then using Eq.(13), we can write
the above equation Eq.(24) as,
\begin{eqnarray}
\Box ^{x}~\frac{\delta w(C)}{\delta \sigma (x)}
=\frac{g^2}{2\pi}~\frac{\delta w(C)}{\delta \sigma (x)}
\end{eqnarray}
This is the basic equation.It is non-local and is valid for
any size of the loop.This is, in fact, an equation for the smeared
gauge invariant field strength.
In abelian $U(1)$ gauge theory, using the anomaly equation
Eq.(2), we obtain
\begin{eqnarray}
\Box ^{x}~\frac{\delta w(C)}{\delta \sigma (x)}
\equiv \partial^{\mu x}~ \partial^{x}_{\mu}~
\frac{\delta w(C)}{\delta \sigma (x)}
=\frac{g^2}{\pi}~\frac{\delta w(C)}{\delta \sigma (x)}
\end{eqnarray}
The equation above for abelian gauge theory should be supplemented
by equation, Eq.(18) , {\it i.e.,}
\begin{eqnarray}
\partial^{x}_{\mu} w(C_{xx}) ~=0  \nonumber\\
\end{eqnarray}
Note that, in the non-abelian gauge theory
there is a local equation for the field strength.The equation is
given by,
\begin{eqnarray}
D^{\mu}D_{\mu} \Phi(x) =\frac{g}{2\pi}~\Phi(x)
\end{eqnarray}
The non-local equation Eq.(25), for the smeared  gauge invariant field
strength is, in fact, obtained by
multiplying both sides of the equation, Eq.(28) by
$w(C_{xx})$, taking trace and using the definitions of area derivative
and the non-local derivative.

\end{document}